# Magnetic excitations in the quantum spin system $NH_4CuCl_3$

Akira Oosawa[1], Toshio Ono[2], Kazuhisa Kakurai[1], Hidekazu Tanaka[3]

[1]Advanced Science Research Center, Japan Atomic Energy Research Institute, Tokai, Ibaraki 319-1195, Japan
[2]Department of Physics, Tokyo Institute of Technology, Oh-okayama, Meguro-ku, Tokyo 152-8551, Japan
[3]Research Center for Low Temperature Physics, Tokyo Institute of Technology, Oh-okayama, Meguro-ku, Tokyo 152-8551, Japan

**Abstract**
$NH_4CuCl_3$ has two magnetization plateaus at one-quarter and three-quarters of the saturation magnetization, irrespective of the external field direction. Magnetic excitations in $NH_4CuCl_3$ have been investigated by means of neutron inelastic scattering experiments. The constant-$\boldsymbol{Q}$ energy scan profiles were collected in $a^*$-$c^*$ plane. Two well-defined magnetic excitations were observed. The dispersion relations along four different directions were determined. Both excitations are found weakly dispersive. The results were compared with those obtained from the ESR measurements.

## 1. Introduction

Quantum spin systems often exhibit numerous peculiar magnetic features, which cannot be interpreted by conventional classical spin models. For instance, a step like magnetization process (magnetization plateau) corresponding to the quantization of the magnetization is the macroscopic quantum phenomenon.

$NH_4CuCl_3$ has the monoclinic structure (space group $P2_1/c$) [1]. In the crystal structure, planar dimers of $Cu_2Cl_6$, in which $Cu^{2+}$ ions have spin-1/2, are stacked on top of one another to form infinite double chains parallel to the crystallographic *a*-axis. These double chains are located at the corners and centre of the unit cell in the *b*-*c* plane, and are separated by $NH_4^+$ ions. The magnetic ground states of the isomorphous compounds $KCuCl_3$ and $TlCuCl_3$ are the spin singlet with excitation gaps [2, 3, 4]. From the analyses of the dispersion relations obtained by neutron inelastic scattering, it was found that the origin of the spin gap is the strong antiferromagnetic interaction in the chemical dimer $Cu_2Cl_6$, and that the neighboring dimers are coupled by the interdimer interactions along the double chain and in the (1, 0, -2) plane. On the other hand, $NH_4CuCl_3$ is a gapless antiferromagnet with $T_N = 1.3$ K [5]. $NH_4CuCl_3$ presents salient magnetization plateaus in the magnetization process at one-quarter and three-quarters of the saturation magnetization [6], *i.e.*, for $H \parallel a$, these plateaus are observed in 5.0 T < $H$ < 12.8 T, and 17.9 T < $H$ < 24.7 T, respectively, and the magnetization saturates at $H_s$ = 29.1 T. Because the magnetization plateaus are observed irrespective of the field direction, the origin of the plateau can be attributed to quantum effect. However, the microscopic mechanism of the magnetization plateau of $NH_4CuCl_3$ cannot be fully comprehended so that further investigations, for instance, the search of magnetic excitations are highly required by using inelastic neutron scattering techniques.

## 2. Experimental details

Neutron inelastic scattering experiment was performed on deuterated $ND_4CuCl_3$ using the C1-1 ISSP-HER spectrometer installed at JRR-3M, Tokai, Japan. It was confirmed that $NH_4CuCl_3$ and $ND_4CuCl_3$ exhibit the same magnetic properties. The constant-$k_f$ mode was taken with a fixed final neutron energy $E_f$ of 4.4 meV. In order to gain intensity, the horizontal focusing analyzer with the applicable collimations was used. A PG-filter and a Be-filter were used to suppress the higher order contaminations. We stacked four single crystals of $ND_4CuCl_3$ with a total volume of approximately 1cm$^3$. The samples were mounted in an ILL-type orange cryostat with it's $a$*- and $c$*-axes in the scattering plane. The crystallographic parameters were determined as $a^* = 1.5663$ Å$^{-1}$, $c^* = 0.68198$ Å$^{-1}$ and $\cos\beta^* = 0.07$ at helium temperatures.

## 3. Results and discussions

Figure 1 shows the constant-$\boldsymbol{Q}$ energy scan profiles in $ND_4CuCl_3$ measured at $T$=1.6 K (> $T_N$) for $\boldsymbol{Q}=(h, 0, 1)$, $(0, 0, l)$, $(h, 0, 2h+1)$ and $(h, 0, -2h+1)$. Two well-defined excitations are observed near $E$ = 1.5 and 3 meV in almost all scans. The scan profiles were fitted with two Gaussians as shown by the solid line in Fig. 1. It was confirmed from the temperature dependence of both excitations that their origin is magnetic. Figure 2 shows the dispersion relations $\omega(\boldsymbol{Q})$ in $ND_4CuCl_3$ for $\boldsymbol{Q}=(h, 0, 1)$, $(0, 0, l)$, $(h, 0, 2h+1)$ and $(h, 0, -2h+1)$. Both magnetic excitations are less dispersive in all directions. The magnetic excitations in $ND_4CuCl_3$ contrast with the dispersive excitations in isostructural spin gap compounds $KCuCl_3$ and $TlCuCl_3$, where the magnetic excitations are most dispersive along $\boldsymbol{Q}=(h, 0, 2h+1)$ and less dispersive along $\boldsymbol{Q}=(h, 0, -2h+1)$ [2, 3, 4]. Since the ground state is gapless in $ND_4CuCl_3$, there are many dimer sites occupied by the spin triplet even at zero field, although they are less than one-quarter of the total dimer sites. The spin triplet excited on the singlet dimer site cannot hop to the occupied site. As the result, the hopping of the excited triplet is strongly suppressed. Thus, it is considered that the gapless ground state is one of the origin of the less dispersive magnetic excitations in $ND_4CuCl_3$.

The magnetic excitations in $NH_4CuCl_3$ have been investigated by ESR measurements [7, 8]. Two singlet-triplet excitations which correspond to the higher edge fields of the one-quarter and three-quarters magnetization plateaus were observed. Their zero field excitation energies of the excitations were estimated as 388 GHz (1.60 meV) and 708 GHz (2.93 meV), respectively, which are the almost same as the excitation energies at $\boldsymbol{Q}=(0, 0, 1)$ observed in the present measurements. Thus, we infer that the two magnetic excitations observed in the present measurements correspond to two singlet-triplet excitations observed by the ESR measurements. Since $NH_4CuCl_3$ undergoes magnetic ordering at zero field, there should exist a gapless excitation, or spin wave excitation. Since the gapless ESR excitation expressed by $\omega/\gamma=H$ was observed, we can expect that one more magnetic excitation exists below 0.5 meV. However, the excitation could not be observed in present measurements. This may be ascribed to the large incoherent scattering by the $NH_4^+$ ions and $Cl^-$ ions below 1.0 meV, as shown in Fig. 1.

## 4. Conclusion

We have presented the results of the neutron inelastic scattering measurements on deuterated $ND_4CuCl_3$ which exhibits the same magnetic properties as $NH_4CuCl_3$. Two well-defined magnetic excitations were observed in the $a^*$-$c^*$ plane. The dispersion relations of both the magnetic excitations in $ND_4CuCl_3$ were determined as shown in Fig. 2. It was found that the magnetic excitations are less dispersive in all directions. Comparing to the ESR measurements, we conclude that two magnetic excitations observed in the present measurements correspond to the singlet-triplet excitations, which are related to the higher edge fields of the one-quarter and third-quarters magnetization plateaus, respectively.

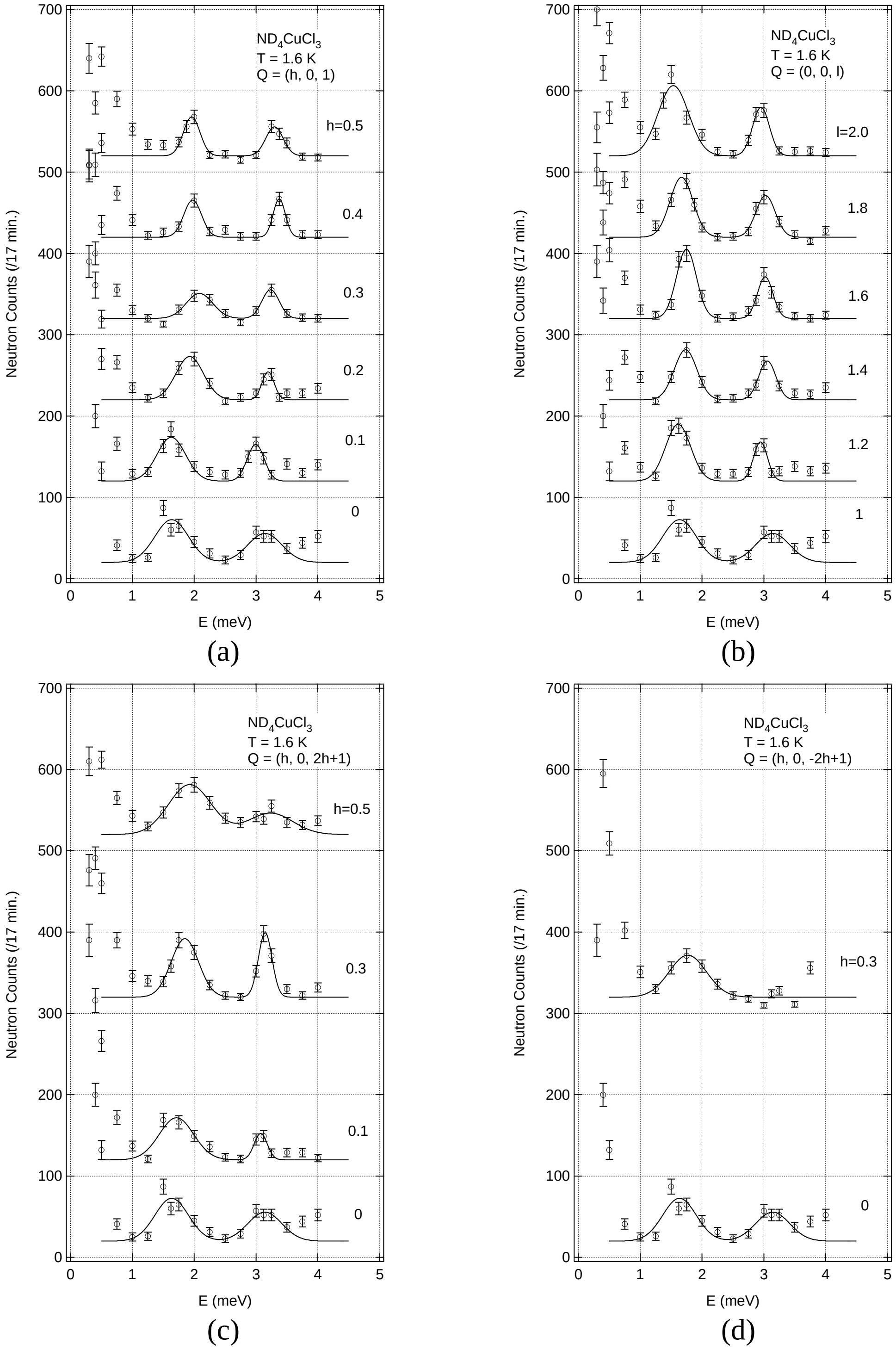

Fig. 1: Profiles of the constant-***Q*** energy scans in $ND_4CuCl_3$ for ***Q*** along (a) ($h$, 0, 1), (b) (0, 0, $l$), (c) ($h$, 0, 2$h$+1) and (d) ($h$, 0, -2$h$+1) with $0 \leq h \leq 0.5$ and $1 \leq l \leq 2$. The solid lines are fits using two Gaussian functions.

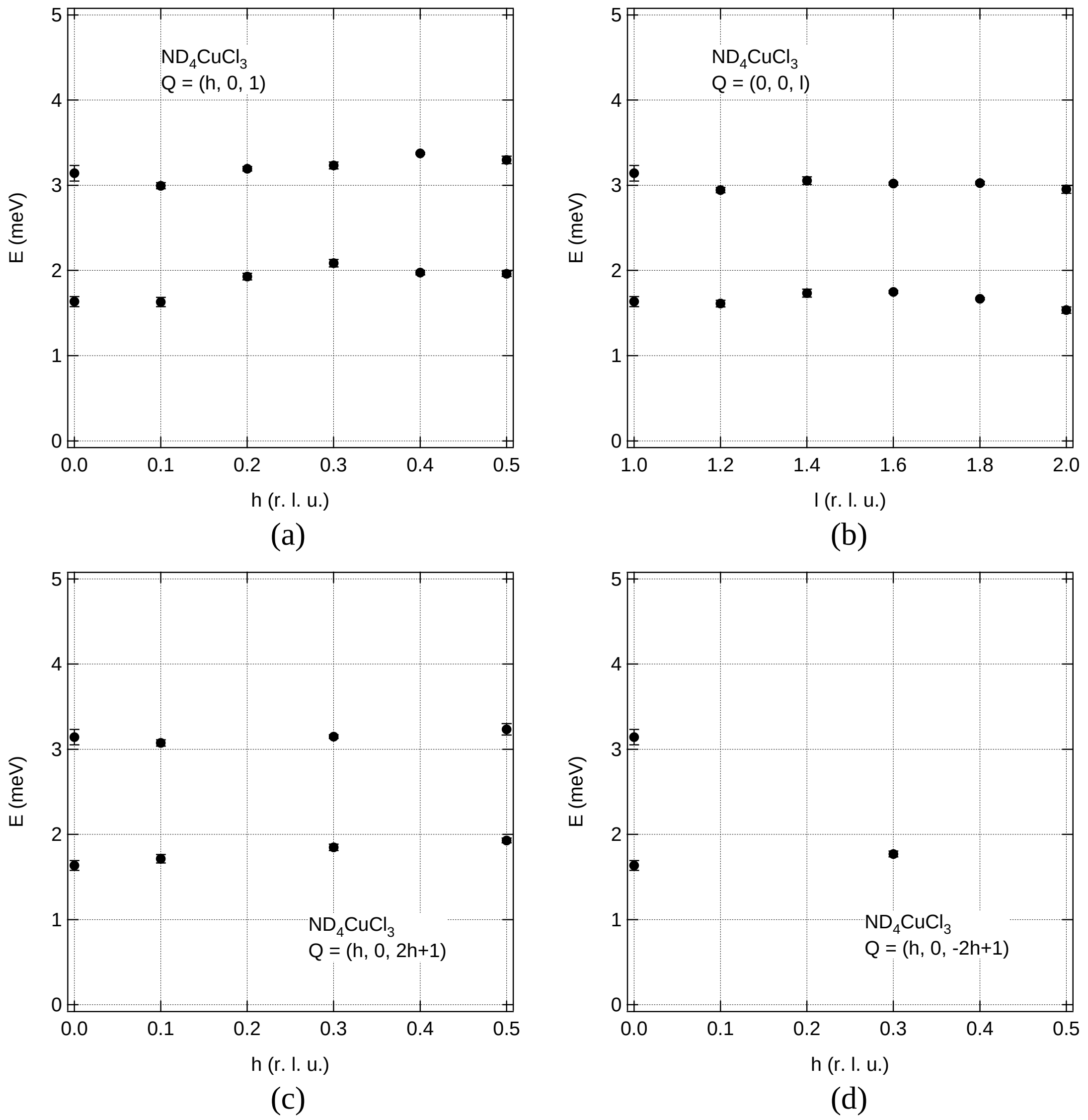

Fig. 2: Dispersion relations $\omega(\boldsymbol{Q})$ in $ND_4CuCl_3$ for $\boldsymbol{Q}$ along (a) ($h$, 0, 1), (b) (0, 0, $l$), (c) ($h$, 0, 2$h$+1) and (d) ($h$, 0, -2$h$+1) with $0 \le h \le 0.5$ and $1 \le l \le 2$.

**Acknowledgments**

This work was supported by the Toray Science Foundation and a Grant-in-Aid for Scientific Research on Priority Areas (B) from the Ministry of Education, Culture, Sports, Science and Technology of Japan.